\documentclass[useAMS,usenatbib,usegraphicx,A4]{mn2e}
\usepackage{aas_macros}
\usepackage[a4paper,top=1.4in, bottom=0.9in, left=0.7in, right=0.6in]{geometry}

\title[Cross-correlation induced by weak lensing]{\textit{Herschel}\thanks{\textit{Herschel} is an ESA space observatory with science instruments provided by European-led Principal Investigator consortia and with important participation from NASA.}\,--ATLAS/GAMA: SDSS cross-correlation induced by weak lensing.}

\author[Gonz\'alez-Nuevo et al.]
{
J. Gonz\'alez-Nuevo,$^{1,2}$\thanks{E-mail: gnuevo@ifca.unican.es} A. Lapi,$^{2,3}$ M. Negrello,$^{4}$ L. Danese,$^{2}$ G. De Zotti,$^{4,2}$
\newauthor
 S. Amber,$^{5}$ M. Baes,$^{6}$ J. Bland-Hawthorn,$^{7}$ N. Bourne,$^{8,9}$ S. Brough,$^{10}$
\newauthor
 R.S. Bussmann,$^{11}$ Z.-Y. Cai,$^{2}$ A. Cooray,$^{12}$ S.P.~Driver,$^{13,14}$ L. Dunne,$^{15}$  S. Dye,$^{8}$ 
\newauthor
 S. Eales,$^{16}$ E. Ibar,$^{17}$ R. Ivison,$^{9}$ J. Liske,$^{18}$ J. Loveday,$^{19}$ S. Maddox,$^{15}$
\newauthor
 M.J. Micha{\l}owski,$^{9}$ A.S.G.~Robotham,$^{13}$ D. Scott,$^{20}$ M.W.L. Smith,$^{16}$
\newauthor
 E. Valiante,$^{16}$ and J.-Q. Xia$^{21}$   
\vspace*{8pt} \\
$^1$ Inst. de Fisica de Cantabria (CSIC-UC), Avda. los Castros s/n, 39005 Santander, Spain\\
$^2$ SISSA, Via Bonomea 265, I-34136 Trieste, Italy\\
$^3$ Dipartimento di Fisica, Universit\`a di Roma `Tor Vergata', Via Ricerca Scientifica 1, 00133 Roma, Italy\\
$^4$ INAF-Osservatorio Astronomico di Padova, Vicolo dell'Osservatorio 5, I-35122 Padova, Italy \\
$^{5}$ Department of Physical Sciences, The Open University, Milton Keynes MK7 6AA, UK \\
$^{6}$ Sterrenkundig Observatorium, Universiteit Gent, Krijgslaan 281 S9, B-9000 Gent, Belgium \\
$^{7}$ Sydney Institute for Astronomy, School of Physics, University of Sydney, NSW 2006, Australia \\
$^{8}$ School of Physics and Astronomy, University of Nottingham, University Park, Nottingham, NG7 2RD, UK \\
$^{9}$ Institute for Astronomy, University of Edinburgh, Royal Observatory, Blackford Hill, Edinburgh EH9 3HJ, UK \\
$^{10}$ Australian Astronomical Observatory, PO Box 915, North Ryde, NSW 1670, Australia \\
$^{11}$ Harvard-Smithsonian Center for Astrophysics, 60 Garden Street, Cambridge, MA 02138, USA \\
$^{12}$ Dept. of Physics \& Astronomy, University of California,Irvine, CA 92697, USA \\
$^{13}$ International Centre for Radio Astronomy Research (ICRAR), University of Western Australia, Crawley, WA 6009, Australia \\
$^{14}$ Scottish Universities' Physics Alliance (SUPA), School of Physics and Astronomy, \\ ~~~University of St Andrews, North Haugh, St Andrews, KY16 9SS, UK\\
$^{15}$ Department of Physics and Astronomy, University of Canterbury, Private Bag 4800, Christchurch, 8140, New Zealand \\
$^{16}$ School of Physics and Astronomy, Cardiff Univ., The Parade, Cardiff, CF24 3AA, UK \\
$^{17}$ Instituto de F\'isica y Astronom\'ia, Universidad de Valpara\'iso, Avda. Gran Breta\~na 1111, Valpara\'iso, Chile \\
$^{18}$ European Southern Observatory, Karl-Schwarzschild-Str. 2, 85748 Garching, Germany \\
$^{19}$ Astronomy Centre, University of Sussex, Falmer, Brighton BN1 9QH, UK \\
$^{20}$ Department of Physics \& Astronomy, University of British Columbia,Vancouver, BC V6T 1Z1, Canada \\
$^{21}$ Key Laboratory of Particle Astrophysics, Institute of High Energy Physics, \\        ~~~Chinese Academy of Science, P.O. Box 918-3, Beijing 100049, China \\
}

\begin{document}

\maketitle
\clearpage
\begin{abstract}
We report a highly significant ($>10\sigma$) spatial correlation between galaxies with $S_{350\mu\rm m}\ge 30\,$mJy detected in the equatorial fields of the \textsl{Herschel} Astrophysical Terahertz Large Area Survey (H-ATLAS) with estimated redshifts $\ga 1.5$, and SDSS or GAMA galaxies at $0.2\le z\le 0.6$. The significance of the cross-correlation is much higher than those reported so far for samples with non-overlapping redshift distributions selected in other wavebands. Extensive, realistic simulations of clustered sub-mm galaxies amplified by foreground structures confirm that the cross-correlation can be explained by weak gravitational lensing ($\mu<2$). The simulations also show that the measured amplitude and range of angular scales of the signal are larger than can be accounted for by galaxy-galaxy weak lensing. However, for scales $\la 2\,$arcmin, the signal can be reproduced if SDSS/GAMA galaxies act as signposts of galaxy groups/clusters with halo masses in the range $10^{13.2}$--$10^{14.5} {\rm M}_{\odot}$. The signal detected on larger scales appears to reflect the clustering of such halos.
\end{abstract}

\begin{keywords}
methods: statistical -- cosmology: observations -- large-scale structure of Universe -- infrared: galaxies.
\end{keywords}

\section{Introduction}

Light rays coming from a distant source are deflected by the foreground gravitational field. This, on the one hand, stretches the area of a given sky region, thus decreasing the surface density of sources and, on the other hand, magnifies the background sources, increasing their chances of being included in a flux-limited sample. The net effect, termed ``magnification bias'', is extensively described in the literature \citep[see, e.g.,][]{Schneider1992}. It implies that an excess/decrease of background sources from a flux-limited sample will be found in the vicinity of matter overdensities \citep{BartelmannSchneider1993,MoessnerJain1998,Scranton2005}. The amplitude of the excess increases with the slope of the background source number counts. Thus, gravitational lensing constitutes a direct probe of the cosmic gravitational field. The most dramatic manifestations of lensing, called ``strong lensing'', which includes multiple images, arcs or ``Einstein rings'', are rare however. These manifestations show up on angular scales of arcseconds and provide information on high density structures such as galaxies or galaxy clusters. The lower density structures, which include most of the mass in the Universe, nevertheless, can still produce observable effects via ``weak lensing''. The magnification bias due to weak lensing modifies the galaxy angular correlation function, because the observed images do not coincide with true source locations \citep{Gunn1967,Kaiser1992,Moessner1998,Loverde2008}, but the effect is generally small and difficult to single out. An unambiguous manifestation of weak lensing is the cross-correlation between two source samples with non-overlapping redshift distributions. The occurrence of such correlations has been tested and established in several contexts: 8$\sigma$ detection of cosmic magnification from the galaxy-quasar cross-correlation function \citep{Scranton2005}; a simultaneous detection of gravitational magnification and dust reddening effects due to galactic haloes and large-scale structure \citep[galaxy-quasar cross-correlation;][]{Menard2010}; and a 7$\sigma$ detection of a cross-correlation signal between $z\sim3-5$ Lyman-break galaxies and \textit{Herschel} sources \citep{Hildebrandt2013} among others. See also \citet{BartelmannSchneider2001} and references therein.

Since gravitational magnification decreases the effective detection limit, the amplitude of the magnification bias increases with increasing steepness of the number counts of background sources, and is then particularly large at sub-mm wavelengths, where the counts are extremely steep \citep{Clements10,Oliver2010}. As demonstrated by \citet{Negrello2010} this property can be used to effectively identify strongly lensed galaxies in the sub-mm band, opening a new era of gravitational lensing studies \citep{Ivison2010,Swinbank10,Conley2011,Cox2011,Bussmann2012,Bussmann2013,Harris2012,Fu2012,Wardlow2013}.
At the same time, for a survey covering a sufficiently large area the counteracting effect on the solid angle is small \citep{JainLima2011}. A substantial fraction of galaxies detected by deep large area \textsl{Herschel} surveys at 250, 350 and $500\,\mu$m with the Spectral and Photometric Imaging Receiver \citep[SPIRE;][]{Griffin10} reside at $z\ga 1.5$ \citep{Amblard2010,Lapi11,Pearson2013} and therefore constitute an excellent background sample for the Sloan Digital Sky Survey (SDSS) galaxies, which are located at $z\la 0.8$ (with a median redshift of $z\sim 0.1$).

A first attempt at measuring lensing-induced cross-correlations between \textsl{Herschel}/SPIRE galaxies and low-$z$ galaxies was carried out by \citet{Wang2011}, who found convincing evidence of the effect. 
The analysis can now be made with much better statistics, thanks to the availability of catalogues of \textsl{Herschel}/SPIRE sources covering much larger areas, new releases of SDSS data and the spectroscopic measurements of the Galaxy And Mass Assembly \citep[GAMA; ][]{Driver2011} with many more galaxy spectra than SDSS for those areas \citep{Baldry2010}. Such an improved study is the subject of this paper. The paper is structured as follows. In Sect.~\ref{sect:samples} we describe the selection of background and foreground samples. In Sect.~\ref{sect:correl} we present our estimates of the auto- and cross-correlation functions, while in Sect.~\ref{sect:sims} we describe the simulations we have carried out to interpret the results. The main results are summarized and discussed in Sect.~\ref{sect:conclusions}. A complementary analysis using cross-correlation measurements with different selections of both foreground and background samples, aimed at constraining the redshift distribution of the background sources, has been carried out by Schneider et al. (in preparation). In addition an analysis of the effect of lensing on the identification of sub-millimetre galaxies in the \textsl{Herschel} Astrophysical Terahertz Large Area Survey \citep[H-ATLAS;][]{Eales10} has been carried out by Bourne et al. (in preparation).

Throughout the paper we adopt a flat $\Lambda \rm CDM$ cosmology with matter density $\Omega_{\rm m} = 0.32$, $\Omega_{\Lambda} = 0.68$ and Hubble constant $h=H_0/100\, \rm km\,s^{-1}\,Mpc^{-1} = 0.67$ \citep{PlanckCollaborationXVI2013}.

\begin{figure}
\includegraphics[width=\columnwidth]{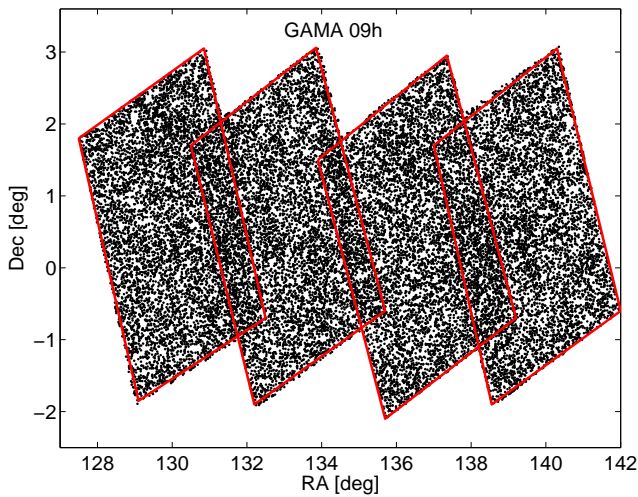}
\caption{Distribution of selected H-ATLAS sources in the 9h equatorial field (one third of the total equatorial regions). The field is made of four partially overlapping areas, referred to as ``tiles''.}
\label{fig:tiles}%
\end{figure}

\section{Sample selection}\label{sect:samples}

We have selected our background sample from the catalogue of sources detected in the three H-ATLAS equatorial fields, covering altogether $\simeq 161\,\hbox{deg}^2$. The H-ATLAS is the largest area extragalactic survey carried out by the \textsl{Herschel} space observatory \citep{Pilbratt10}, covering $\sim 550\,\hbox{deg}^2$ with PACS \citep{Poglitsch10} and SPIRE \citep{Griffin10} instruments between 100 and $500\,\mu$m. Details of the H-ATLAS map-making, source extraction and catalogue generation can be found in \citet{Ibar10}, \citet{Pascale11}, \citet{Rigby11}, Maddox et al. (in preparation) and Valiante et al. (in preparation).

\begin{figure}
\begin{center}
\includegraphics[width=\columnwidth]{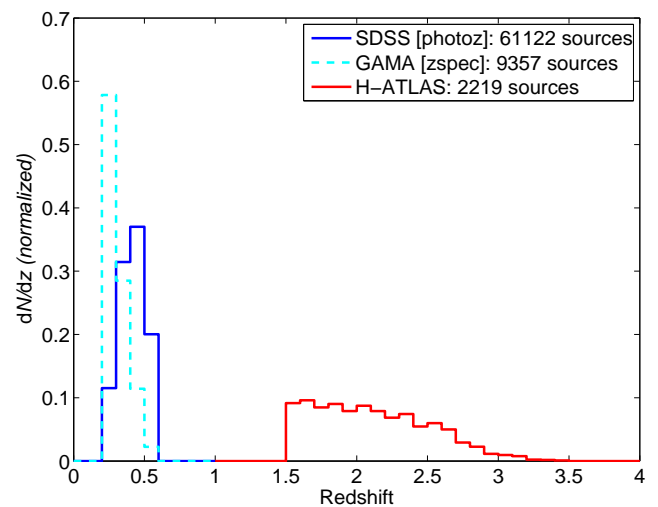}
\caption{Redshift distributions of the background H-ATLAS sample (red histogram), the SDSS photometric sample (in blue), and the GAMA spectroscopic one (in cyan) for one of the tiles.}
\label{fig:xwth_zdists}%
\end{center}
\end{figure}

To extract high-redshift ($z\ge 1.5$) candidate galaxies we have adopted, similar to \citet{Lapi11}, the following criteria: (i) $250\,\mu$m flux density $S_{250}\ge 35\,$mJy; (ii) no Sloan Digital Sky Survey (SDSS) counterpart with reliability $R>0.8$, as determined by \citet{Smith11}; and (iii) $\ge 3\sigma$ detection at $350\,\mu$m. We added the further constraint $S_{350}\ge 30\,$mJy to facilitate comparison with the simulations described in Sect.~\ref{sect:sims}. As pointed out by \citet{Lapi11} requirement (ii) introduces a bias against H-ATLAS sources that are strongly lensed by SDSS galaxies closely aligned with them (which may be misinterpreted as their optical counterparts). Such objects are rare, however, and their loss is irrelevant for the purposes of this paper.

Next we have estimated the photometric redshifts of the selected galaxies by means of  a minimum $\chi^2$ fit of a spectral energy distribution (SED) template to the PACS (which are mostly upper limits) and SPIRE data. As shown by \citet{Lapi11} and \citet{gnuevo12}, a good template is the SED of SMM~J2135$-0102$ \citep[``The Cosmic Eyelash'' at $z=2.3$;][]{Ivison2010,Swinbank10}. A comparison with spectroscopic redshifts has shown that the use of this template does not introduce any systematic offset and has reasonably low rms error \citep[median $\Delta z/(1+z)\equiv (z_{\rm phot}-z_{\rm spec})/(1+z_{\rm spec})=-0.002$, with a dispersion of 0.115 and no outliers;][]{gnuevo12}. Similar conclusions were obtained by \citet{Pearson2013}, confirming our approach. Our background sample comprises all sources with photometric redshift $z\ge 1.5$, $26{,}630$ sources in total. The distribution of selected sources on the plane of the sky for the 9h H-ATLAS fields (one third of the total equatorial survey) is shown, as an example, in Fig.~\ref{fig:tiles}. The area covered is made of four partially overlapping regions (hereafter ``tiles''), each of about $3.8^\circ\times 3.8^\circ$. The estimated redshift distribution of selected sources in one of the tiles is shown in Fig.~\ref{fig:xwth_zdists}. Similar redshift distributions are found with the other tiles. The accuracy of our photometric redshift estimates is enough to avoid any overlap with foreground sources, for which we require $z\le 0.6$.

Our main foreground sample is extracted from the Ninth Data Release (DR9) of the SDSS \citep{Ahn2012}\footnote{http://www.sdss3.org/dr9.} and comprises all galaxies in the H-ATLAS fields with $r<22$ and \textit{photometric} redshifts in the range $0.2\le z\le 0.6$ (hereafter the \textit{photoz} sample). Following the results of \citet{Lapi12}, we do not expect a relevant gravitational lensing effect for galaxies at $z<0.2$, while their much higher number would slow down our calculations. On the other hand, the number of galaxies at $z>0.6$ is so relatively small that they can not produce any important statistical contribution to our results. We prefer to increase the redshift gap between the foreground-background samples in order to minimize the potential cross-contamination (see Sect. \ref{sect:cross-cont}).
We did not impose any constraint on the quality of the photometric redshift, to avoid biasing the sample towards lower redshift galaxies that are less effective as gravitational lenses. Taking into account that galaxies with low stellar masses, $M_\star\la 10^{10.5} {\rm M}_\odot$, produce negligible amplifications for the angular scales considered in our analysis, we have arbitrarily set a lower limit of $M_\star>10^{10.4} {\rm M}_\odot$ (see Sect.~\ref{sect:sims} for more details about mass estimation). In addition, we have also imposed an upper limit to the $r$-band luminosity, $L_r<10^{11.6} \rm{L}_\odot$, in order to remove sources with anomalously bright $r$-band luminosities, which are probably due to overestimates of the photometric redshifts ($\la 0.3 \%$ of the total). At about this limit we have noticed an excess of galaxies with respect to the luminosity function derived by \citet{Ber13} (see Sect. \ref{sect:sims}). The sample comprises $\sim 5\times10^4\,$galaxies per tile, i.e. $686{,}333$ galaxies in total. The redshift distribution for one of the tiles is shown by the blue histogram in Fig.~\ref{fig:xwth_zdists}. The median value is $z_{\rm phot, med}= 0.42$.

\begin{figure}
\begin{center}
\includegraphics[width=\columnwidth]{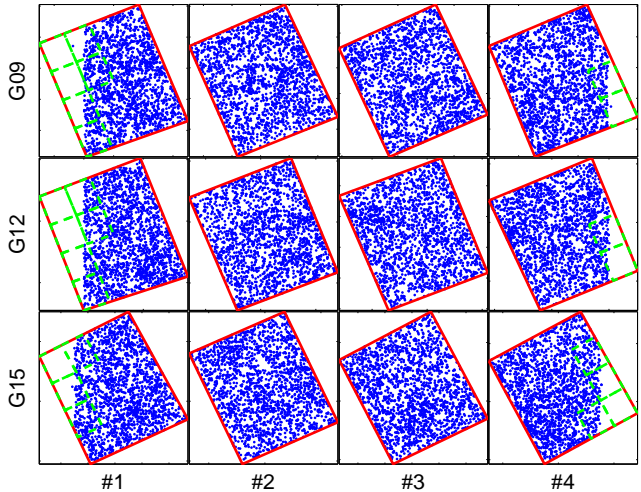}
\caption{Distribution of the \textit{zspec} sample sources in the 12 tiles. The 24 mini-tiles with green boundaries have obvious coverage ``holes'' and were not used in the analysis. For clarity, only the first 2000 randomly selected galaxies per tile are shown.}
\label{fig:zspec_tiles}%
\end{center}
\end{figure}

\begin{figure}
\begin{center}
\includegraphics[width=\columnwidth]{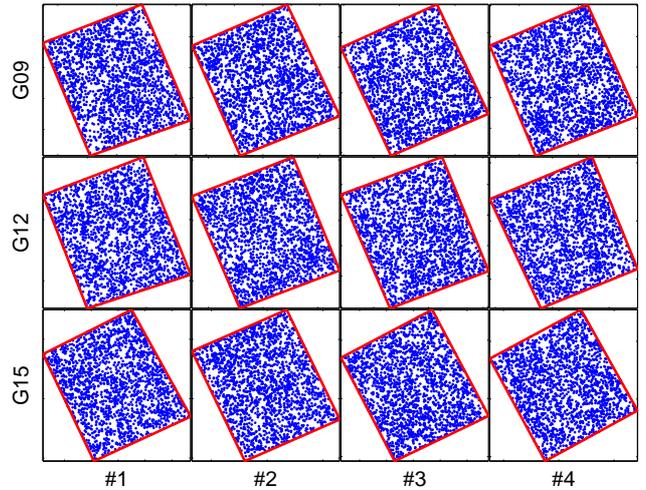}
\caption{Distribution of the \textit{photoz} sample sources in the 12 tiles. For clarity, only the first 2000 randomly selected galaxies per tile are shown.}
\label{fig:photoz_tiles}%
\end{center}
\end{figure}

Accurate redshift measurements of the foreground sources are important to carry out realistic simulations of the effect of the gravitational lensing, which is related to the relative angular diameter distances between the observer, the lens and the source. 
On account of that, we have also defined a spectroscopic sample (hereafter the \textit{zspec} sample) drawn from the GAMA II spectroscopic survey \citep[][Liske et al., in prep]{Driver2009}. As a result of the coordination of the two surveys, there is substantial overlap among survey regions of H-ATLAS and GAMA. In particular, both surveys observed three equatorial regions at 9, 12 and 14.5 h (referred to as G09, G12 and G15, respectively). The GAMA II equatorial survey regions are $12^{\circ}\times5^{\circ}$ in size, and were surveyed down to a limit of $r < 19.8$ mag. For our \textit{zspec} sample we select all GAMA II galaxies (from TilingCatv40) with reliable redshift measurements, which were made with a new fully automatic redshift code \citep{Baldry2014}, and $0.2 < z < 0.6$.
This sample is smaller than the \textit{photoz} one. It comprises $\simeq 9000\,$galaxies per tile, i.e. $101{,}514\,$galaxies in total. Their median redshift, $z_{\rm spec, med}= 0.28$ is significantly lower than for the \textit{photoz} sample, as shown by the cyan histogram in Fig.~\ref{fig:xwth_zdists}. Note that the magnification is far less sensitive to errors in the photometric redshifts of background sources, $\sigma_z \simeq 0.115 (1+z)$, since they have a small effect on the ratios of observer--source/lens--source angular diameter distance ratios.

A check of the distribution of galaxies in the \textit{zspec} sample has shown that its coverage does not exactly match the H-ATLAS one, as illustrated in Fig.~\ref{fig:zspec_tiles}. To cure this, as well as to minimize the possible sample variance effect, we have divided each tile into 16 equal area mini-tiles (with green boundaries shown in Fig.~\ref{fig:zspec_tiles}) and we have excluded from further analysis the 24 mini-tiles with only a partial coverage. The \textit{photoz} sample is immune to this problem, as illustrated in Fig.~\ref{fig:photoz_tiles}.

\section{Correlation functions}\label{sect:correl}

\subsection{Autocorrelation functions}

\begin{figure}
\begin{center}
\includegraphics[width=\columnwidth]{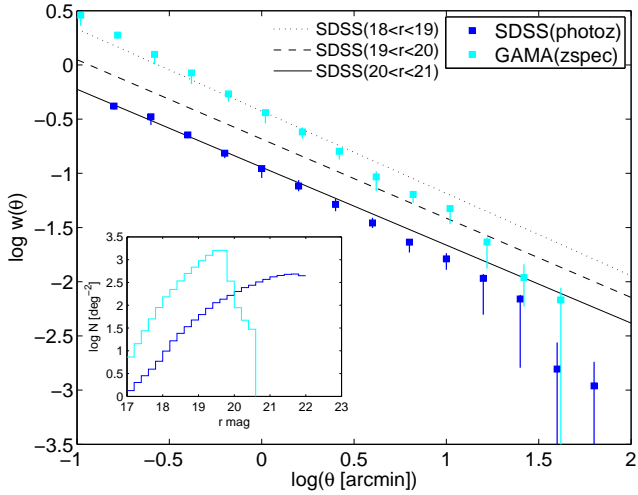}
\caption{Autocorrelation function of our foreground samples compared with determinations of the best fit power laws using the full SDSS catalogue, split into  magnitude intervals \citep{Con02,Wan13}. The inset shows the distributions of $r$-band magnitudes for the two samples. }
\label{fig:sdss_wth}%
\end{center}
\end{figure}

\begin{figure}
\begin{center}
\includegraphics[width=\columnwidth]{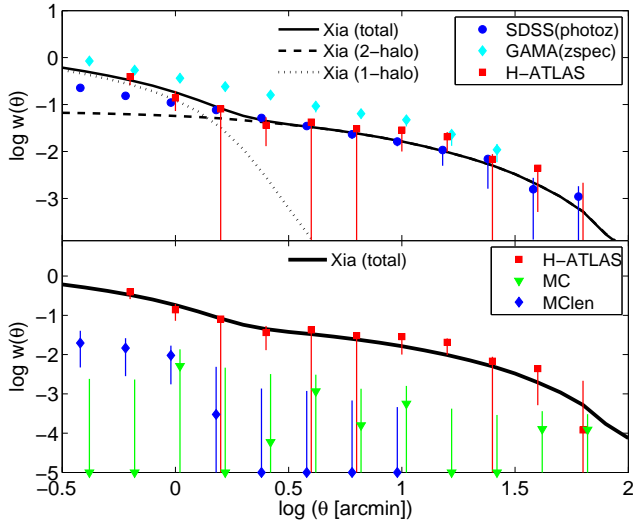}
\caption{\textit{Top panel}: Autocorrelation functions of the H-ATLAS sample (red squares) and of the SDSS ones (blue squares, photometric; cyan squares, spectroscopic). Black lines show the expectations for the \citet{Xia12} model for the background sources (dotted line, 1-halo term; dashed line, 2-halo term; solid line, total).  \textit{Bottom panel}: measured autocorrelation function of the H-ATLAS sample (red squares) compared with the expected contribution from weak lensing using the \textit{zspec} sample and assuming that galaxies are signposts of groups of galaxies/cluster halos (blue diamonds), based on $\sim4500$ Monte Carlo simulations (see text for more details). The green upper limits come from $\sim4500$ Monte Carlo simulations of random distributions. The black solid line is the same as in the upper panel.}
\label{fig:autocor}%
\end{center}
\end{figure}

The angular (auto)-correlation function, $w(\theta)$, is a measure of the probability, in excess of the expectation for a Poisson distribution, of finding a galaxy within each of two infinitesimal solid angles 
separated by an angle $\theta$, $P(\theta) = N[1 + w(\theta)]$, where $N$ is the mean surface density of galaxies. We have computed $w(\theta)$ for the background and foreground samples using the \citet{LandySzalay1993} estimator
\begin{equation}
w(\theta)={{\rm DD}(\theta)-2{\rm DR}(\theta)+{\rm RR}(\theta)\over {\rm RR}(\theta)},
\end{equation}
where DD$(\theta)$ is the number of unique pairs of real sources with separation $\theta$, DR$(\theta)$ is the number of unique pairs between the real catalogue and a mock sample of sources with random positions, and RR$(\theta)$ is the number of unique pairs in the random source catalogue.

We computed the $w(\theta)$ of SDSS/GAMA galaxies for each tile (in order to determine it on scales of up to $\sim 1^\circ$, not possible with the mini-tiles) and plotted in Fig.~\ref{fig:sdss_wth} the median values and the associated uncertainties estimated using the ``median statistics'' of \citet{Got01}. 
Median statistics lead to results that are quite similar and almost as constraining as $\chi^2$ likelihood methods, but with somewhat more robustness since they do not assume Gaussianity of the errors or that their magnitudes are known.
If we make $N$ independent measurements $M_i$, where $i=1,\ldots, N$, the probability that exactly $n$ of the $N$ measurements are higher than the true median is given by the binomial distribution, $P=2^{-N}N!/[n!(N-n)!]$. Thus, if the measurements are ranked by value, such that $M_j>M_i$ for $j>i$, then the probability that the true median lies between $M_i$ and $M_{i+1}$ is
\begin{equation}
P=\frac{2^{-N}N!}{[i!(N-i)!]},
\end{equation}
therefore providing a method for obtaining the confidence limits.

Our estimate for the photometric catalogue has low uncertainties and is in line with previous determinations carried out for several intervals of $r$-band magnitudes using the full SDSS catalogue \citep{Con02,Wan13}, taking into account that our sample is dominated by sources with $r\sim21$. As shown in the inset of Fig. \ref{fig:sdss_wth}, the \textit{zspec} sample is more weighted towards brighter sources ($r\leq19$) due to an obvious selection bias. In agreement with SDSS auto-correlation results, the (magnitude limited) \textit{zspec} sample is more clustered than the \textit{photoz} one.

The angular (auto)-correlation function of our background sample is shown in Fig.~\ref{fig:autocor} (red filled squares). 
The computed auto-correlation was limited to angular scales $\ga30$ arcsecs to avoid the potential bias caused by the resolution of the instruments \citep[FWHM$\sim18$ and 25 arcsec for the 250 and $350\mu$m bands, respectively;][]{Rigby11}. In the top panel of Fig.~\ref{fig:autocor} the $w(\theta)$ of the background sources is compared with those of the foreground samples, as well as with the prediction of the \citet{Xia12} best fit model. 

This model investigated the clustering of submillimetre galaxies in a self-regulated baryon collapse scenario by combining the physical evolutionary model for proto-spheroid galaxies of \citet{Granato04} and an independent halo occupation distribution analysis. The cosmic infrared background power spectra at wavelengths from $250 \mu\rm{m}$ to $2 \rm{mm}$ measured by \textit{Planck}, \textit{Herschel}, South Pole Telescope and Atacama Cosmology Telescope experiments  \citep[see references in][]{Xia12} were fitted using the halo model with only two free parameters: the minimum halo mass; and the power-law index of the mean occupation function of satellite galaxies.
The prediction shown in Fig.~\ref{fig:autocor} (black lines) was computed adopting the redshift distribution of Fig.~\ref{fig:xwth_zdists}, without any adjustment of the parameters. 

The consistency between the model and the data is excellent. The signal is clearly detected up to scales $\ga 50$\,arcmin; it is dominated by the 2-halo term on scales above $\simeq 2$\,arcmin and by the 1-halo term on smaller scales. The lower panel will be commented upon in Sect.~\ref{sect:sims}.

\subsection{Cross-correlation functions}

\begin{figure}
\begin{center}
\includegraphics[width=\columnwidth]{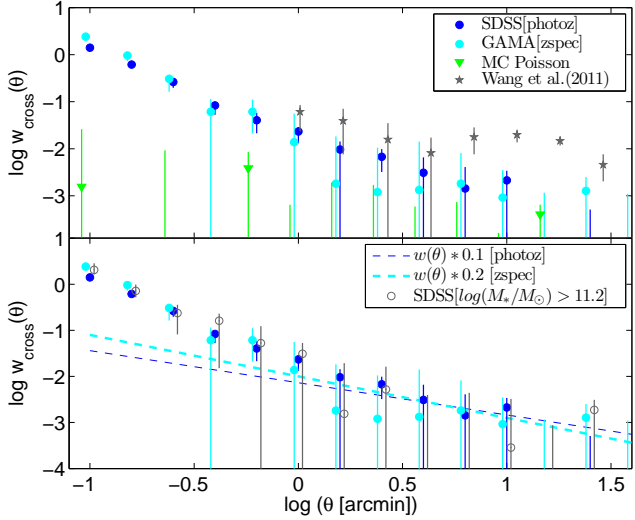}
\caption{\textit{Top panel}: Cross-correlation between the H-ATLAS sample and the photometric and spectroscopic foreground samples (blue and cyan circles, respectively). Measurements from \citet{Wang2011} are also plotted for comparison (grey stars). The green upper limits are the result of $\sim2000$ realistic Monte Carlo simulations, not accounting for the effect of lensing. Such simulations were made to check our procedure and to further illustrate the significance of the detected cross-correlation signal. \textit{Bottom panel}: comparison of the measured cross-correlations with the auto-correlation functions of the photometric and spectroscopic foreground samples, scaled down by a factor of $\sim10$ (blue dashed line) and $\sim20$ (cyan dashed lines), respectively. The cross-correlation between the H-ATLAS sample and the photometric foreground sample, limited to $\log(M_\star/\rm{M}_\odot)>11.2$, is also shown.}
\label{fig:xwth}%
\end{center}
\end{figure}

The cross-correlation function of two population of sources is the fractional excess probability, relative to a random distribution, of finding a source of population 1 and a source of population 2, respectively, within infinitesimal solid angles
separated by an angle $\theta$ \citep{Peebles1980}. We have computed the cross-correlation between our background and foreground samples using a modified version of the \citet{LandySzalay1993} estimator \citep{Her01}
\begin{equation}
w_{\rm cross}(\theta)=\frac{\rm{D}_1\rm{D}_2-\rm{D}_1\rm{R}_2-\rm{D}_2\rm{R}_1+\rm{R}_1\rm{R}_2}{\rm{R}_1\rm{R}_2}
\end{equation}
where $\rm{D}_1\rm{D}_2$, $\rm{D}_1\rm{R}_2$, $\rm{D}_2\rm{R}_1$ and $\rm{R}_1\rm{R}_2$ are the normalized data1-data2, data1-random2, data2-random1 and random1-random2 pair counts for a given separation $\theta$ \citep[see][for a discussion of different estimators of $w_{\rm cross}(\theta)$]{Blake2006}.

\begin{figure}
\begin{center}
\includegraphics[width=\columnwidth]{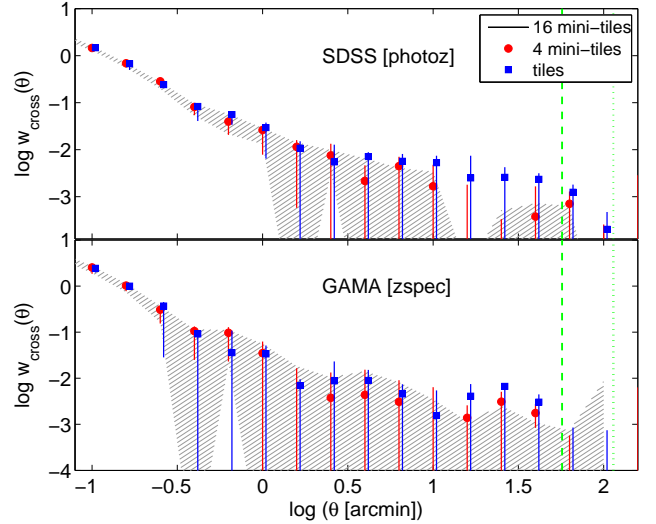}
\caption{Observed cross-correlation between the H-ATLAS sample and the \textit{photoz}(\textit{top panel}) and \textit{zspec}(\textit{bottom panel}) samples (grey hatched area; dividing each tile in 16 mini-tiles), compared with the measured ones dividing each tile into 4 mini-tile (red circles) or using the whole tiles (blue squares). The green dashed and dotted lines indicate the typical sizes of the 16 and 4 mini-tiles, respectively.}
\label{fig:xwth_tiles}%
\end{center}
\end{figure}

We have followed the same procedure as in the auto-correlation case by computing the angular cross-correlation function, but this time for each of the 192 mini-tiles, and estimating the median values and the associated uncertainties using the ``median statistics'' of \citet{Got01}. Use of this procedure attempts to minimize the sample variance effect.

The measured angular cross-correlations between the foreground and background samples are shown in Fig.~\ref{fig:xwth}. Unlike the auto-correlation case, the small angular scale limit is dictated by the H-ATLAS positional error (relatively to which the SDSS error is negligibly small) whose rms value at $250\,\mu$m is $\sim2.4\,$arcsec for $5\,\sigma$ sources \citep{Rigby11,Smith11}. To be conservative, we have limited our computations to angular scales $\ga 6$ arcsec, greater than twice the H-ATLAS positional error. Our results are in broad agreement with those of \citet{Wang2011}, although we do not confirm the strong signal reported by them on scales of tens of arcmin. The median values, uncertainties and additional statistical information on the cross-correlation results for both foreground samples can be found in Tables~\ref{tab:xwth_stats} and \ref{tab:xwth_stats_zspec}, where we also give the probability, from the Kolmogorov-Smirnov test, that the measured median $w_{\rm cross}(\theta)$ is obtained from a random source distribution. It is clear from the table that the significance of the signal from the \textit{photoz} sample is very high. This is in fact the best detection of lensing-induced cross-correlation reported so far.

\begin{table}
\begin{minipage}{85mm}
\caption{Measured cross-correlation between the H-ATLAS background sample and the photometric foreground galaxies, compared with the random source distribution case (estimated from $\sim2000$ Monte Carlo simulations).\label{tab:xwth_stats}}
\begin{center}
\begin{tabular}{rrrr}

\hline%
$\log\theta$ & signal & random & $\log p$%
\footnote{Here $p$ is the probability, from the Kolmogorov-Smirnov test, that the measured median $w_{\rm cross}(\theta)$ is obtained from the random source distribution.}\\
~[arcmin] & [$\pm$95\% CL] & [$\pm$95\% CL] & \\
\hline%
$-$1.0 & 1.4400{\tiny $_{-0.1500}^{+0.2200}$} & $-$0.0197{\tiny $_{-0.0440}^{+0.0302}$} & $-$ 77.9 \\ [1ex] 
$-$0.8 & 0.6480{\tiny $_{-0.1090}^{+0.0530}$} & 0.0112{\tiny $_{-0.0238}^{+0.0165}$} & $-$ 62.6 \\ [1ex] 
$-$0.6 & 0.2420{\tiny $_{-0.0440}^{+0.0400}$} & $-$0.0072{\tiny $_{-0.0113}^{+0.0116}$} & $-$ 33.1 \\ [1ex] 
$-$0.4 & 0.0758{\tiny $_{-0.0328}^{+0.0195}$} & $-$0.0027{\tiny $_{-0.0061}^{+0.0070}$} & $-$  8.2 \\ [1ex] 
$-$0.2 & 0.0348{\tiny $_{-0.0124}^{+0.0177}$} & $-$0.0033{\tiny $_{-0.0039}^{+0.0044}$} & $-$  7.0 \\ [1ex] 
 0.0 & 0.0236{\tiny $_{-0.0098}^{+0.0092}$} & 0.0013{\tiny $_{-0.0038}^{+0.0032}$} & $-$  4.8 \\ [1ex] 
 0.2 & 0.0067{\tiny $_{-0.0083}^{+0.0076}$} & $-$0.0010{\tiny $_{-0.0021}^{+0.0019}$} & $-$  2.4 \\ [1ex] 
 0.4 & 0.0061{\tiny $_{-0.0025}^{+0.0042}$} & 0.0003{\tiny $_{-0.0015}^{+0.0012}$} & $-$  3.4 \\ [1ex] 
 0.6 & 0.0021{\tiny $_{-0.0049}^{+0.0045}$} & $-$0.0001{\tiny $_{-0.0008}^{+0.0008}$} & $-$  4.1 \\ [1ex] 
 0.8 & 0.0017{\tiny $_{-0.0030}^{+0.0013}$} & $-$0.0001{\tiny $_{-0.0008}^{+0.0005}$} & $-$  3.8 \\ [1ex] 
 1.0 & 0.0016{\tiny $_{-0.0022}^{+0.0015}$} & 0.0000{\tiny $_{-0.0004}^{+0.0003}$} & $-$  5.6 \\ [1ex] 
 1.2 & -0.0013{\tiny $_{-0.0013}^{+0.0013}$} & $-$0.0003{\tiny $_{-0.0002}^{+0.0002}$} & $-$  2.4 \\ [1ex] 
 1.4 & -0.0010{\tiny $_{-0.0011}^{+0.0016}$} & 0.0001{\tiny $_{-0.0001}^{+0.0001}$} & $-$  4.8 \\ [1ex] 
\hline%
\end{tabular}
\end{center}
\end{minipage}
\end{table}

The limited number of foreground galaxies slightly diminishes the significance of the signal derived from the \textit{zspec} sample. However, the similarity between the signal from both foreground samples makes us confident that potential systematic uncertainties in the photometric redshift estimation are not an issue.

Below $20--30\,$arcmin the measured cross-correlation is almost independent of the mini-tile size used (see Fig.~\ref{fig:xwth_tiles}). At larger angular scales the observed signal, using 16 or 4 mini-tiles per tile, are biased low. However, we decided to continue dividing the tiles in 16 smaller areas due to the lower uncertainties at angular scales below $\sim3\,$arcmin.

The bottom panel of Fig.~\ref{fig:xwth} compares the measured cross-correlations with the auto-correlation functions of the photometric and spectroscopic foreground samples, scaled down by factors of $\sim10$ and $\sim20$, respectively. These factors correspond to the probability of lensing with moderate amplifications, $\mu<2$, for the different median redshifts of the samples \citep{Lapi12}. This comparison suggests that, above a few arcmin, the cross-correlation functions reflect the clustering properties of the foreground samples. We have investigated this possibility and, more generally, the interpretation of the cross-correlation signals by means of extensive simulations, described in Sect.\,\ref{sect:sims}.

\begin{table}
\begin{minipage}{85mm}
\caption{Measured cross-correlation between the H-ATLAS background sample and the spectroscopic foreground galaxies, compared with the random source distribution case (estimated from $\sim2000$ Monte Carlo simulations).\label{tab:xwth_stats_zspec}}
\begin{center}
\begin{tabular}{rrrr}
\hline%
$\log\theta$ & signal & random & $\log p$%
\footnote{Here $p$ is the probability, from the Kolmogorov-Smirnov test, that the measured median $w_{\rm cross}(\theta)$ is obtained from the random source distribution.}\\
~[arcmin] & [$\pm$95\% CL] & [$\pm$95\% CL] & \\
\hline%
$-$1.0 & 2.4050{\tiny $_{-0.4850}^{+0.3650}$} &  0.0000{\tiny $_{-0.0001}^{+0.0001}$} & $-$ 37.9 \\ [1ex] 
$-$0.8 & 0.9585{\tiny $_{-0.2335}^{+0.1715}$} & $-$0.0755{\tiny $_{-0.0170}^{+0.0166}$} & $-$ 25.2 \\ [1ex] 
$-$0.6 & 0.3095{\tiny $_{-0.1315}^{+0.0735}$} & $-$0.0189{\tiny $_{-0.0101}^{+0.0090}$} & $-$  8.9 \\ [1ex] 
$-$0.4 & 0.0655{\tiny $_{-0.0630}^{+0.0311}$} & $-$0.0085{\tiny $_{-0.0073}^{+0.0065}$} & $-$  3.3 \\ [1ex] 
$-$0.2 & 0.0641{\tiny $_{-0.0485}^{+0.0459}$} & $-$0.0029{\tiny $_{-0.0047}^{+0.0038}$} & $-$  2.9 \\ [1ex] 
 0.0 & 0.0163{\tiny $_{-0.0341}^{+0.0394}$} & 0.0008{\tiny $_{-0.0029}^{+0.0019}$} & $-$  1.9 \\ [1ex] 
 0.2 & 0.0014{\tiny $_{-0.0174}^{+0.0151}$} & $-$0.0004{\tiny $_{-0.0018}^{+0.0020}$} & $-$  0.1 \\ [1ex] 
 0.4 & $-$0.0010{\tiny $_{-0.0092}^{+0.0097}$} & $-$0.0005{\tiny $_{-0.0010}^{+0.0011}$} & $-$  0.6 \\ [1ex] 
 0.6 & 0.0012{\tiny $_{-0.0078}^{+0.0107}$} & $-$0.0008{\tiny $_{-0.0006}^{+0.0006}$} & $-$  2.3 \\ [1ex] 
 0.8 & 0.0023{\tiny $_{-0.0071}^{+0.0067}$} & 0.0001{\tiny $_{-0.0004}^{+0.0004}$} & $-$  2.5 \\ [1ex] 
 1.0 & 0.0004{\tiny $_{-0.0055}^{+0.0029}$} & 0.0000{\tiny $_{-0.0003}^{+0.0003}$} & $-$  3.0 \\ [1ex] 
 1.2 & $-$0.0006{\tiny $_{-0.0033}^{+0.0019}$} & $-$0.0000{\tiny $_{-0.0002}^{+0.0002}$} & $-$  1.1 \\ [1ex] 
 1.4 & 0.0011{\tiny $_{-0.0022}^{+0.0016}$} & $-$0.0000{\tiny $_{-0.0001}^{+0.0001}$} & $-$  2.7 \\ [1ex]
\hline%
\end{tabular}
\end{center}
\end{minipage}
\end{table}

Finally, we have checked that most of the cross-correlation signal for the \textit{photoz} sample is produced by the most massive SDSS sources, $\log(M_\star/\rm{M}_\odot)>11.2$ or $\log(M_h/\rm{M}_\odot)\ga13.2$  (Fig. \ref{fig:xwth}, grey circles). However, the lower number of such objects translates into a larger uncertainty in the measured signal.

\subsection{Assessment of the potential cross-contamination}\label{sect:cross-cont}

\begin{figure}
\begin{center}
\includegraphics[width=\columnwidth]{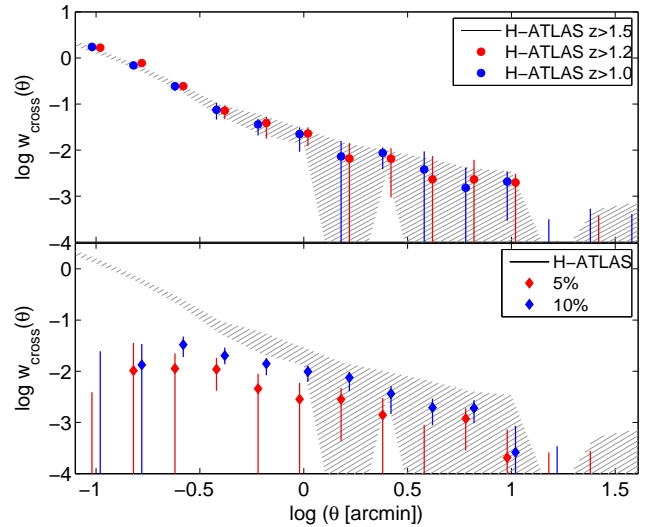}
\caption{\textit{Top panel}: Estimated cross-correlation between the \textit{photoz} sample and a background sample that extends to lower redshifts (red circles, $1.2<z<4.0$; blue circles, $1.0<z<4.0$). \textit{Bottom panel}: Potential cross-contamination inferred from simulations if 5 or 10\% of the background sources are considered as foreground galaxies (randomly selected in each case). In both panels, the grey hatched area corresponds to the observed cross-correlation between the H-ATLAS sample and the \textit{photoz} sample.}
\label{fig:xcont}%
\end{center}
\end{figure}

As discussed in detail in \citet{Lapi11}, a “cold” observed SED may be associated either with a low-$z$ cirrus-dominated galaxy or to a redshifted warm galaxy, introducing a potential degeneracy. This problem can be overcome, however, because cold, low-$z$ galaxies are only moderately obscured by dust (the cirrus optical depth cannot be very large), and are therefore relatively bright in the optical bands. For this reason, as described in Sect.\,\ref{sect:samples}, we removed from the background sample those sources that have an SDSS counterpart with reliability $R>0.8$, even if this requirement introduces a bias against H-ATLAS strongly lensed sources. For the rest of the background sources, so far we have assumed that errors on photometric redshifts do not cause any overlap between background and foreground samples. 

H-ATLAS sources with intermediate values of reliability, $0.1<R<0.8$, (i.e. that have a non-negligible probability of being SDSS/GAMA sources that we could have incorrectly considered as H-ATLAS sources at $z>1.5$), constitute only $\sim9\%$ of the background sample. In addition, the limiting colour used to separate elliptical from spiral galaxies \citep[$g-r>0.8$;][]{Ber10} increases with redshift, but remains approximately constant around $g-r\sim1.5$ in the redshift range $0.2<z<0.6$, for typical elliptical SEDs. We have checked that $\sim 30\%$ of these sources have colours compatible with elliptical galaxies, $g-r>1.5$, and therefore are expected to have sub-mm emission that is hardly detectable by \textit{Herschel}. 

For the other $\sim70$\%, following \citet{gnuevo12}, we have verified that the ratios between the $r$-band flux density and those at 350 and $500\,\mu$m are much higher than expected for spiral galaxies, or even for the median local starburst SED \citep{Smith2012}, both placed at $z=0.6$. In fact only $\sim1\%$ of those sources, for the spiral SED, and $\sim6\%$, for the local starburst one, show compatible ratios. This means that the SDSS magnitudes of those sources are too faint to account for the optical and far-IR emissions at the same time, \textit{if they have the cold far-IR SEDs observed for $z<0.6$ galaxies with moderate star formation rates}. In other words, most of the remaining background sources have higher apparent far-IR to optical luminosity ratios than the spiral or \citet{Smith2012} galaxies, akin to those of sub-mm galaxies, and/or have colder far-IR colours. This implies that they must be at higher redshifts than those indicated by the optical/near-IR SEDs of their SDSS counterparts.
Therefore, from this photometric analysis, we estimated that the fraction of SDSS/GAMA sources that we could have incorrectly considered as H-ATLAS sources at $z>1.5$ is lower than $0.09\times0.6\times0.06\sim0.4\%$ (for the median local starburst SED case).

A similar upper-limit, $\sim1\%$, is obtained if we consider the removed $R>0.8$ sources as a rough indicator of the cross-contamination (remember that many strongly lensed galaxies fall in this category). The fraction of background sources with an $R>0.8$ optical counterpart in the foreground sample is roughly $9\%$. Of these sources, again only $\sim9\%$ have apparent far-IR to optical luminosity ratios compatible with local starburst galaxies at $z<0.6$.

A small part of the H-ATLAS area is covered by a very deep VLA survey of the Subaru Deep Field (Ivison et al. in preparation). We have used this survey to find very reliable ($R>0.99$) radio counterparts for approximately 100 H-ATLAS sources. These sources have been preliminarily identified in deep optical/near-IR imaging (from the Subaru Deep Field) and high-quality photometric redshifts have been derived from these data. Of the 27 sources that satisfy the parent sample selection criteria (see Sect.\,\ref{sect:samples}), 24 are at $z>0.9$. Therefore, there is a maximum potential cross-contamination of $\sim10\%$, although probably much lower, because some of these three particular cases could be foreground galaxies acting as lenses (Allen et al. in preparation).

The H-ATLAS parent sample selection procedure produces a Gaussian-like redshift distribution peaked around $z\sim2$, with a lower tail that extends to $z\sim0.5$ \citep[see figure 4 of][]{Lapi11}. By selecting only those sources with $z>1.5$ we are minimizing the potential cross-contamination and maintaining the bulk of the sample at the same time. However, we can relax this minimum redshift in order to verify that, if an overlap exists, its cross-contamination effect in the measured cross-correlation signal is negligible. As shown in Fig.~\ref{fig:xcont} (\textit{top panel}), the measured signal remains almost the same, independent of the lower redshift limit used, hence confirming the non-overlap assumption.

Finally, we have used the simulations described in Sect.\,\ref{sect:sims}, to assess the potential cross-contamination produced by a non-negligible fraction of mismatched background versus foreground sources (see Fig.~\ref{fig:xcont}, \textit{bottom panel}). In this particular set of simulations, a fraction of background sources are randomly selected and moved to the position of randomly selected foreground ones. 
The results show that a $\sim 10\%$ contamination by overlapping background/foreground sources is ruled out, because it would yield  a far stronger cross-correlation signal on scales $\ge 1$~arcmin than measured for our samples, yet falling short of accounting for the observed signal on smaller scales.

\section{Realistic simulations}\label{sect:sims}

In this paper we focus on the measured cross-correlation signal between two different samples, without taking into account possible completeness issues or selection biases in the foreground samples. This decision, aimed at maximizing the number of sources in both samples, complicates the theoretical interpretation of the measured signal. For this reason, we carried out realistic simulations of the background sources, using the actual information (position, photometry and redshift) for the foreground sources. With this approach the potential statistical issues with the foreground sources are fully dealt with.

\subsection{Simulation set-up and results}
We have generated simulated 3-dimensional distributions of background sources down to $S_{350}=10\,$mJy, drawing them from the redshift-dependent luminosity functions of un-lensed galaxies yielded by the \citet{Cai2013} model \citep[which updates the model by][and reproduces the observed counts, after allowing for the effect of gravitational lensing]{Lapi11}. Since the cross-correlations are induced mainly by weak lensing, a flux density limit of 10\,mJy is sufficient for the parent sample, given that the flux density limit of the background sample we want to simulate is 30\,mJy. 
The clustering properties of the sources were generated using the software described in \citet{gnuevo2005}, with the spatial correlation functions $\xi(r,z)$ given by the \citet{Xia12} model. This model successfully reproduces the clustering properties of \textsl{Herschel} galaxies, as well as the power spectra of fluctuations of the cosmic infrared background measured by both \textsl{Planck} and \textsl{Herschel}. We performed typically 10 simulations per mini-tile.

\begin{figure}
\begin{center}
\includegraphics[width=\columnwidth]{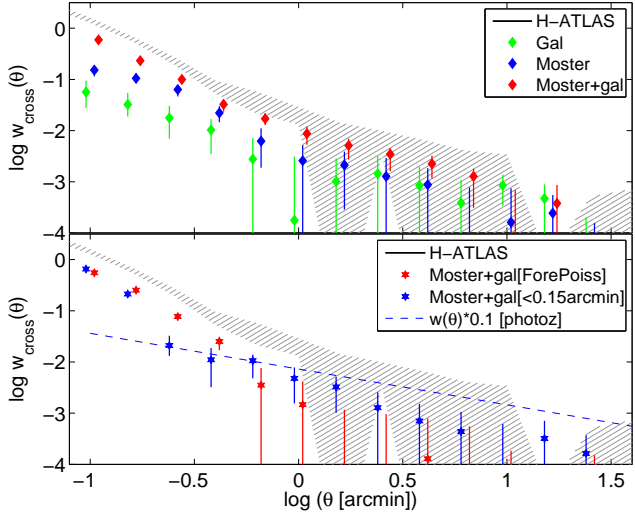}
\caption{\textit{Top panel}: Observed cross-correlation between the H-ATLAS sample and the \textit{photoz} samples (grey hatched area) compared with the one inferred from simulations of lensing by galactic halos (green diamonds), by group/cluster halos with NFW density profiles (magenta diamonds), and a combination of both (red diamonds). \textit{Bottom panel}: Effect of ignoring the clustering properties of foreground galaxies is illustrated by simulations of the effect of lensing by a random distribution of foreground galaxies (``Moster+gal'' case; red stars). On the other hand, by limiting the direct effect of lensing to angular scales $<0.15$ arcmin, the recovered cross-correlation signal reflects the auto-correlation of the foreground galaxies (blue stars) above such scales. The blue dashed line shows the power-law representation of the auto-correlation of foreground sample, scaled down by a factor of 10. The grey hatched area shows the same observed cross-correlation as in the top panel.}
\label{fig:xwth_sim}%
\end{center}
\end{figure}

\begin{figure}
\begin{center}
\includegraphics[width=\columnwidth]{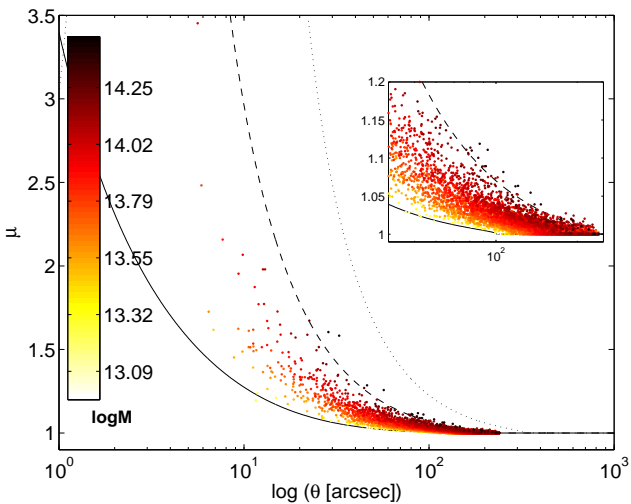}
\caption{Gravitational amplifications of background sources as a function of angular distance and lens halo mass (see the colour scale on the left) obtained in the simulations of a single tile. The black lines correspond to a source at redshift $z_{\rm s}=2.5$ and lenses with $M_{\rm h}=3\times 10^{13}$, $3\times 10^{14}$ and $10^{15}\,{\rm M}_\odot$ (solid, dashed and dotted lines, respectively), all at $z_{\rm lens}=0.5$.}
\label{fig:mu_theta_Mh}%
\end{center}
\end{figure}

Next, we have estimated the magnification of each background source by the foreground source closest to its line of sight, using the formalism of \citet{Lapi12} that requires four ingredients: $z_{\rm lens}$; $z_{\rm source}$; halo mass of the lens, $M_{h}$; and angular separation between source and lens (triaxiality effects were neglected). The calculations were performed for several choices of the relation between $M_{\rm h}$ and the rest-frame $r$-band luminosity of the lens, using both the NFW \citep{Navarro1996,Lokas2001} and classic singular isothermal sphere (SIS) density profiles. The flux densities of the background sources were updated, taking the gravitational amplification into account, and the simulated background sample was constructed by selecting sources with $S_{350}>30\,$mJy and $z\ge 1.5$. For the foreground samples we used either the real positions and redshifts of sources or we randomized the positions, keeping the measured or estimated redshifts the same.

As a preliminary test we computed the cross-correlation between the simulated background sample and the foreground samples, ignoring the effect of lensing, in the same way as we did for real samples. As illustrated by the green points in the upper panel of Fig.~\ref{fig:xwth}, no significant correlation was found.

In the next set of simulations we estimated $M_{\rm h}$ using the relationship with the $r$-band luminosity, $L_r$, derived by \citet{Shankar2006} for elliptical galaxies (galaxy-galaxy lensing). In this case, the chosen density profile of the foreground sources was the SIS one, and we allowed for strong-lensing. Two conclusions are immediately apparent (see green diamonds in the top panel of Fig.~\ref{fig:xwth_sim}).
\begin{enumerate}
\item A statistically significant cross-correlation appears on small scales, demonstrating that the observed signal is indeed due to gravitational lensing.

\item The cross-correlation due to galaxy-galaxy lensing falls rapidly on scales larger than $\sim30\,$arcsec. Even on smaller scales it has an insufficient amplitude; therefore it cannot account for the detected signal.
\end{enumerate}
The reason for the second result is illustrated in Fig.~\ref{fig:mu_theta_Mh}, showing that the gravitational amplification by galaxy-scale lenses is negligible on scales larger than $40-50\,$arcsec. Much more massive lenses, which also give stronger effects, must be advocated. This suggests that galaxies in the foreground samples act as signposts for much more massive halos, i.e. are typically the central galaxy of a group or cluster of galaxies.

\begin{figure}
\begin{center}
\includegraphics[width=\columnwidth]{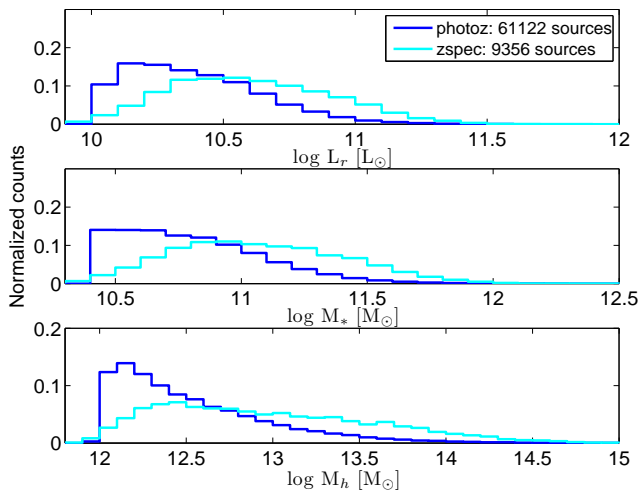}
\caption{Normalized distribution of the estimated $r-$band luminosities, $L_r$, stellar masses, $M_\star$, and halo masses, $M_{\rm h}$, of both foreground samples in a typical tile (see text for more details).}
\label{fig:L_Mh_dists}%
\end{center}
\end{figure}

\begin{figure}
\begin{center}
\includegraphics[width=\columnwidth]{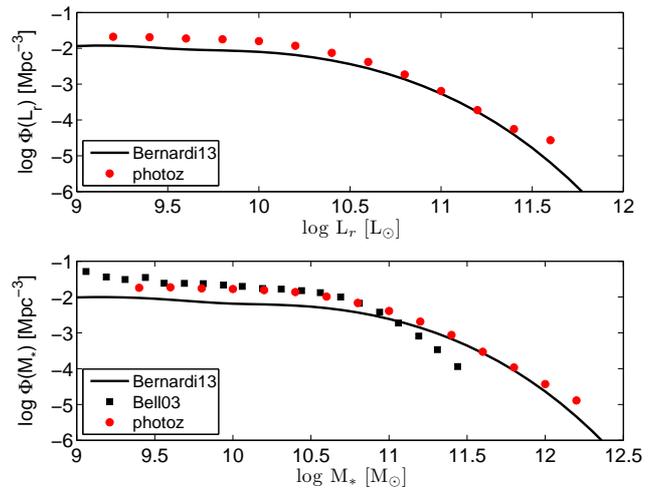}
\caption{Luminosity function (\textit{top}) and stellar mass function (\textit{bottom}) estimated from the \textit{photoz} sample (red circles). The luminosity function derived by \citet[][black line, \textit{top panel}]{Ber13} and the stellar mass functions from \citet[][black line, \textit{bottom panel}]{Ber13} and \citet[][black squares, \textit{bottom panel}]{Bell2003} are shown for comparison.}
\label{fig:LF_MF_check}%
\end{center}
\end{figure}

To test this hypothesis we have associated each galaxy with a halo mass, $M_{\rm h}$, estimated in the following way. From the $r$-band luminosity, $L_r$, we estimated the stellar masses, $M_{\star}$, using a modified version of the luminosity--stellar mass relationship of \citet{Ber03} and \citet{Shankar2006}. We used a lower $(M/L)_0$ value than the original one by \citet{Ber03}, derived for the typical $(g-r)=0.8$ elliptical colours, and we introduced an evolution correction factor \citep[see][ for more details]{Bell2003,Ber10}:
\begin{equation}
M_\star/L_r=3 \times (L_r/10^{10.31})^{0.15}\cdot10^{-0.19z},
\end{equation}
with $M_\star$ and $L_r$ in $\rm{M}_\odot$ and $\rm{L}_\odot$, respectively. Then we used the relationship between the stellar mass and halo mass from  \citet[][their equations 2 and 23--26 with the redshift-dependent parameters from table 7]{Moster2010}, which was determined imposing consistency between both mass functions. The normalized distribution of the estimated $L_r$, $M_\star$ and $M_{\rm h}$ for both foreground samples are shown in Fig.~\ref{fig:L_Mh_dists} for a typical tile. The luminosity, stellar mass and halo mass functions derived from these distributions are in agreement with the latest results \citep[see Fig.~\ref{fig:LF_MF_check},][ and references therein]{Moster2010,Ber13}. 

The upper panel of Fig.~\ref{fig:xwth_sim} shows the cross-correlation signal recovered by considering the weak lensing effect (we limited our analysis to angular separations $>5$ arcsec) produced by massive halos with an NFW density profile. One might worry that associating such a massive halo with every foreground galaxy may be counting the same halo multiple times. Such events must be rare, however, because the typical comoving distance between a galaxy in the \textit{photoz} sample and its nearest neighbour are $\sim 6.5\,$Mpc. Although the recovered signal is stronger than in the galaxy-galaxy lensing case, there is still a lack of power, probably indicating that our assumptions to build the simulations are not complex enough to completely reproduce the measured signal.

There are essentially two main ways to increase the number of background sources that appear, in projection, very close to the foreground ones, i.e. to increase the strength of the cross-correlation signal. Firstly, we can adopt higher halo masses and/or higher numbers of deflectors, or satellites, per halo. However, increasing halo masses would lead to a density of massive halos in excess of that obtained from N-body simulations. The second alternative makes reference to the fact that we have adopted the rather extreme assumption that each dark matter halo has either an NFW or an SIS density profile centred at the position of the SDSS sources, without any sub-structure. Therefore, in order to understand the possible role of halo sub-structures, we have slightly modified the simulation set up in the following way. We associate an SIS profile to each SDSS source with mass estimated from the \citet{Shankar2006} relationship. The SIS profile is centred on the SDSS galaxy. Then we place an NFW mass profile 5 arcsec away from the SDSS source position, in a random direction, and estimate its mass using the \citet{Moster2010} relationship, as explained above. In other words, we have now assumed that the SDSS galaxies act as signposts of much more massive halos, but are not located exactly at their centres.

The signal recovered using this set of simulations, \textit{Moster+gal} (see Fig.~\ref{fig:xwth_sim}), is in better agreement with the measured one. As expected from the halo model formalism \citep[see e.g.][]{Cooray2002}, this last result confirms that sub-structures within a halo, i.e. satellites, have an important role to play in explaining the measured signal. Our simple approach is not able to completely reproduce the signal on scales below $\sim20{-}30\,$arcsec. However, more sophisticated simulations, beyond the scope of the current paper will be performed in a future work.

The direct effect of gravitational lensing by massive halos (the 1-halo term) is not enough to account for the observed signal on arcmin scales. This was shown by repeating the simulations for the case of random distributions of both the background and foreground samples, so that any intrinsic auto-correlation is washed out. As a result (red stars in the bottom panel of Fig.~\ref{fig:xwth_sim}) the statistical significance of the cross-correlations becomes marginal on scales larger than $\simeq 0.5\,$arcmin. This proves that the signal on larger scales is due to the clustering of the large halos \citep[2-halo term; e.g.][]{Cooray2002,Mandelbaum2013}, i.e. it comes from the higher probability that background sources are amplified above the adopted flux density limit, $30\,$mJy, in the neighbourhood of the foreground deflectors. This interpretation is supported by the change in the slope of the cross-correlation function for $\theta \ge 1\,$arcmin. It appears to reflect the auto-correlation function of the foreground sample, sketched in Fig.~\ref{fig:xwth_sim} by the blue dashed line (corresponding to its power-law representation with amplitude decreased by a factor of 10). This factor is probably related to the probability of lensing; for the typical properties of foreground and background samples the factor is around $\sim10$\% for $\mu\sim1.5$ \citep{Lapi12}. In fact, if we take into account the effect of lensing only for separations between background and foreground sources of less than $9\,$arcsec (blue stars in the bottom panel of Fig.~\ref{fig:xwth_sim}), allowing for the possibility of strong lensing in this case, the recovered cross-correlation signal tends to reflect the auto-correlation of the foreground galaxies.

\subsection{Further considerations}\label{sect:plus}

We conclude that the observed cross-correlation is the combination of two contributions. On sub-arcmin scales it comes directly from weak lensing of group/cluster sized halos, while on larger scales it reflects the clustering of such halos. 

This conclusion implies that the observed spatial distribution of the background sources is influenced by the clustering of foreground galaxies. The extreme case would be that the clustering properties of the background sample were completely due to the lensing effect. Therefore we are led to ask, how much of the H-ATLAS sources clustering is intrinsic and not contributed by weak lensing effects?

To answer this question we have performed a set of 100 simulations per mini-tile, using the \textit{zspec} foreground sample and intrinsically unclustered background galaxies. After lensing, the population indeed shows significant clustering, but at a level well below the observed one, $<4\%$ (see Fig.~\ref{fig:autocor}, bottom panel). Strictly speaking, the lensing-induced clustering inferred from simulations is a lower limit to the real measurement, because simulations necessarily take into account only a subset of foreground galaxies. We expect, however, that the bulk of the effect comes from the redshift range covered by our foreground samples.

\begin{figure}
\begin{center}
\includegraphics[width=\columnwidth]{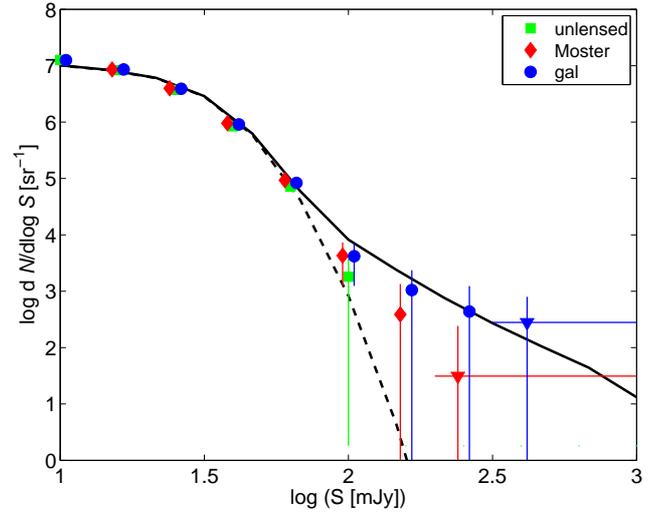}
\caption{Gravitational lensing effect on the recovered source number counts at $350\,\mu$m from the Monte Carlo simulations (using the \textit{zspec} sample). The black lines show the number counts from the \citet{Lapi11} model, with (solid) and without (dashed) including the effect of gravitational lensing. The corresponding simulated counts are represented by the blue circles and by the green squares: the latter represent the unlensed counts, the former include the effect of galaxy-galaxy lensing. The red diamonds show the effect on the counts of weak lensing by group/cluster halos. The triangle symbols at the bright end are upper limits.}
\label{fig:sim_counts}%
\end{center}
\end{figure}

Previous measurements of the counts of sub­millimetre galaxies have been corrected for the effects of galaxy­-galaxy lensing
 \citep{Perrotta02,Perrotta03,Negrello2007,Negrello2010,Paciga2009,HezavehHolder2011,Wardlow2013}. However, the fact that the cross-correlation is dominated by lensing from group/cluster halos suggests that substantial contributions may also come from these more massive halos. As illustrated in Fig.~\ref{fig:sim_counts} our simulations confirm that galaxy-galaxy strong lensing indeed dominates the effect on the counts, but a substantial contribution, at the brightest flux densities, may also come from much larger halos.

As in the previous case, the lensing-induced clustering inferred from simulations is a lower limit to the real one, because simulations do not include all the foreground galaxies. 

The AzTEC millimetre survey of the COSMOS field \citep{Austermann2009,Aretxaga2011} showed an observed excess in number counts at 1.1-mm and a spatial correlation with the optical-IR galaxy density, which seem consistent with lensing of a background submillimetre galaxy population by foreground mass structures along the line of sight. Our measured cross-correlation signal, and, in particular, the weak lensing effect produced by foreground structures on the bright end of the source number counts, appear to confirm these findings.

\section{Conclusions}\label{sect:conclusions}

We have reported a highly significant correlation between the spatial distribution of H-ATLAS galaxies with estimated redshift $\ga 1.5$ and that of SDSS/GAMA galaxies at $0.2\le z\le 0.6$. Extensive, realistic simulations have shown that the cross-correlation is explained by weak gravitational lensing. The much higher significance compared to those reported so far is a result of the extreme steepness of the sub-mm source counts.

The simulations also show that the amplitude and the range of angular scales of the signal are larger than can be accounted for by galaxy-galaxy weak lensing. However, it can be reproduced, on sub-arcmin scales, if SDSS/GAMA galaxies act as signposts of galaxy groups/clusters with halo masses in the range $\sim 10^{13.2}{-}10^{14.5} \rm{M}_\odot$. The signal detected on larger scales appears to reflect the clustering of such halos. Future work will try to extract quantitative astrophysical information about the dark matter halos by comparing the measured cross-correlation signal with dedicated simulations, considering the distribution of sub-halos and their densities, as well as observational constraints like angular resolution and sensitivity.

We have also investigated the effect of clustering of foreground galaxies on the observed angular correlation function of H-ATLAS galaxies. We find that lensing can indeed induce an apparent clustering of randomly distributed background galaxies, but the amplitude of the corresponding angular correlation function is at least a factor of 10 lower than observed.

Finally we find that, although halos of group/cluster size are the dominant contributors to the cross-correlation between H-ATLAS and SDSS/GAMA galaxies, the gravitational magnification effects on counts of sub-mm sources are nevertheless dominated by galaxy-galaxy strong lensing.

\section*{Acknowledgments}
I thank Michael D. Schneider for useful suggestions and comments on the manuscript. The work has been supported in part by the Spanish Ministerio de Ciencia e Innovacion, AYA2012-39475-C02-01, and Consolider-Ingenio 2010, CSD2010-00064, projects and by ASI/INAF Agreement I/072/09/0 for the {\it Planck} LFI activity of Phase E2. JGN acknowledges financial support from the Spanish CSIC for a JAE-DOC fellowship, co-funded by the European Social Fund. MN acknowledges final support from PRIN INAF 2012 project ``Looking into the dust-oscured phase of galaxy formation through cosmic zoom lenses in the\textsl{Herschel} Astrophysical Large Area Survey''. The \textit{Herschel}-ATLAS is a project with \textit{Herschel}, which is an ESA space observatory, with science instruments provided by European-led Principal Investigator consortia and with important participation from NASA. The H-ATLAS website is http://www.h-atlas.org/ . GAMA is a joint European-Australasian project based around a spectroscopic campaign using the Anglo-Australian Telescope. The GAMA input catalogue is based on data taken from the Sloan Digital Sky Survey and the UKIRT Infrared Deep Sky Survey. Complementary imaging of the GAMA regions is being obtained by a number of independent survey programs including \textit{GALEX} MIS, VST KiDS, VISTA VIKING, \textit{WISE}, \textit{Herschel}-ATLAS, GMRT and ASKAP, providing UV to radio coverage. GAMA is funded by the STFC (UK), the ARC (Australia), the AAO, and the participating institutions. The GAMA website is: http://www.gama-survey.org/. This paper has used data from the SDSS-III. Funding for SDSS-III has been provided by the Alfred P. Sloan Foundation, the Participating Institutions, the National Science Foundation, and the U.S. Department of Energy Office of Science. The SDSS-III web site is http://www.sdss3.org/. SDSS-III is managed by the Astrophysical Research Consortium for the Participating Institutions of the SDSS-III Collaboration including the University of Arizona, the Brazilian Participation Group, Brookhaven National Laboratory, University of Cambridge, Carnegie Mellon University, University of Florida, the French Participation Group, the German Participation Group, Harvard University, the Instituto de Astrofisica de Canarias, the Michigan State/Notre Dame/JINA Participation Group, Johns Hopkins University, Lawrence Berkeley National Laboratory, Max Planck Institute for Astrophysics, Max Planck Institute for Extraterrestrial Physics, New Mexico State University, New York University, Ohio State University, Pennsylvania State University, University of Portsmouth, Princeton University, the Spanish Participation Group, University of Tokyo, University of Utah, Vanderbilt University, University of Virginia, University of Washington, and Yale University.

\bibliographystyle{mn2e}
\bibliography{xcorr}

\end{document}